%% file: main.tex
\documentclass{sigchi}

\toappear{
}

\usepackage{balance}       
\usepackage{graphics}      
\usepackage[T1]{fontenc}   
\usepackage{txfonts}
\usepackage{mathptmx}
\usepackage[pdflang={en-US},pdftex]{hyperref}
\usepackage{color}
\usepackage{booktabs}
\usepackage{textcomp}
\usepackage{csvsimple}
\usepackage{floatrow}
\newfloatcommand{capbtabbox}{table}[][\FBwidth]
\usepackage{tabularx}

\usepackage{microtype}        
\usepackage{ccicons}          

\usepackage{todonotes}

\def\plaintitle{The Effectiveness of Haptic Properties \\ Under Cognitive Load: An Exploratory Study}

\def\emptyauthor{}
\def\plainkeywords{Cognitive Load; Haptics; Vibrotactile Feedback; Haptic Interfaces; Haptic Guidelines; Empirical Study;}

\makeatletter
\def\url@leostyle{%
  \@ifundefined{selectfont}{
    \def\UrlFont{\sf}
  }{
    \def\UrlFont{\small\bf\ttfamily}
  }}
\makeatother
\def\pprw{8.5in}
\def\pprh{11in}

\definecolor{linkColor}{RGB}{6,125,233}
\hypersetup{%
  pdftitle={\plaintitle},
  pdfauthor={\emptyauthor},
  pdfkeywords={\plainkeywords},
  pdfdisplaydoctitle=true, 
  bookmarksnumbered,
  pdfstartview={FitH},
  colorlinks,
  citecolor=black,
  filecolor=black,
  linkcolor=black,
  urlcolor=linkColor,
  breaklinks=true,
  hypertexnames=false
}

\begin{document}
\graphicspath{ {./figures/} }

\title{\plaintitle}

\author{%
  \alignauthor{Nava Haghighi\dag, Nathalie Vladis\ddag, Yuanbo Liu\dag, Arvind Satyanarayan\dag\\
    \affaddr{\dag Massachusetts Institute of Technology, \ddag Harvard University}\\
    \affaddr{Cambridge, USA}\\
    \email{\{nava, yuanbo, arvindsatya\}@mit.edu, nathalie\_vladis@hms.harvard.edu}}\\
}


\maketitle

\input{sections/abstract.tex}



\input{sections/introduction.tex}

\input{sections/related-work.tex}

\input{sections/device.tex}

\input{sections/method.tex}

\input{sections/results.tex}

\input{sections/discussion.tex}

\input{sections/acknowledgements.tex}

\balance{}

\bibliographystyle{SIGCHI-Reference-Format}
\bibliography{hapticref}

\end{document}

%% file: sections/abstract.tex
\begin{abstract}
With the rise of wearables, haptic interfaces are increasingly favored to communicate information in an ambient manner. Despite this expectation, existing guidelines are often developed in studies where the participant’s focus is entirely on the haptic task. In this work, we systematically study the cognitive load imposed by properties of a haptic signal. Participants wear a haptic device on their forearm, and are asked to perform a 1-back task. Each experimental condition isolates an individual property of the haptic signal (e.g., amplitude, waveform, rhythm) and participants are asked to identify the gradient of the data. We evaluate each condition across 16 participants, measuring participants’ response times, error rates, and qualitative and quantitative surveys (e.g., NASA TLX). Our results indicate that gender and language differences may impact preference for some properties, that participants prefer properties that can be rapidly identified, and that amplitude imposes the lowest cognitive load.

\end{abstract}

%% file: sections/introduction.tex
\section{Introduction}

With the increasing adoption of wearable devices, there has been a renewed interest in haptics as an alternative to traditional graphical user interfaces (GUIs). 
This interest is primarily because haptic signals can be perceived in an ambient, discrete, and passive manner~\cite{maclean2009putting} and with a smaller response time compared to visual and auditory modalities~\cite{scott2008comparison}. 
Additionally, haptics have been shown to be a robust alternative to visual communication methods when the user’s visual channels are under high cognitive load~\cite{enriquez2008design}. 
Thus far, much of the work on haptics has been focused on low-level perceptual studies such as defining haptic properties \cite{brewster2004tactons}, studying the perceptual limitations such as just noticeable differences (JND) of those properties \cite{gunther2001skinscape, pongrac2008vibrotactile} on perception and mapping of haptic patterns \cite{azadi2014evaluating,lee2012evaluation}, or on meaning association of haptic patterns \cite{maclean2008foundations, seifi2019personalizing, seifi2016exploring}. 
Despite the widely-held expectation that haptics are effective \emph{ambient} interfaces, prior empirical work often examines haptics in isolated environments where the participant’s attention is focused entirely on the haptic task.
There is a gap in systematically studying how these haptic properties and patterns are interpreted ambiently under cognitive load. 

In this paper, we explore five haptic properties\,---\,amplitude, waveform, duration, rhythm, and spatio-temporal pattern\,---\,first introduced by Brewster et al.~\cite{brewster2004tactons}, and evaluate how effectively they encode information under cognitive load.
We conduct a study with 16 participants. 
Participants perform a primary delayed response task (the \emph{1-back} task~\cite{mehler2011agelab}) and are asked to simultaneously interpret a haptic signal, communicated via a haptic feedback device on their forearm.
The haptic signal encodes data using only one of the five properties, and we determine the degree to which it interferes with the primary task by measuring the time taken by participants to respond to the 1-back task, and their error rate for both tasks.
We also administer qualitative and quantitative post-study surveys (e.g., the NASA-TLX~\cite{hart2006nasa, hart1988development}) to gather participants' subjective experiences and preferences.

Our results show that the \textit{Amplitude} condition, followed by \textit{Duration} had the least cognitive load, and the best performance score on both the n-back and the haptic task. This result aligned with the participant preference in the post-experiment survey as well as the NASA-TLX survey. \textit{Waveform} followed by \textit{Rhythm} imposed the highest cognitive load, and \textit{Rhythm} had the worst performance score on both the n-back and the haptic task. However, the NASA-TLX and participant preference showed that \textit{Waveform} had the highest mental demand and was the least preferred by the participants as well. 

%% file: sections/related-work.tex
\section{Related Work}

Chan et al.~\cite{chan2008designing} identify four factors when designing haptic signals: how easily stimuli can be associated with the target meaning, how easily discernible an item is in a set, the salience of an individual stimulus, and whether that saliency persists under cognitive workloads. 
Prior work in haptics has primarily focused on the first three components. 
For instance, studies have been conducted to determine the perceptual limitations of haptic properties including identifying the just noticeable differences (JND) in amplitude, frequency, and rhythm~\cite{pongrac2008vibrotactile, summers1997information, gunther2001skinscape}, the thresholds for identification~\cite{6548433}, discrimination~\cite{israr2006frequency} and resolution~\cite{meier2015exploring}, and the impact of body location~\cite{azadi2014evaluating}.

Researchers have begun to codify these experimental results into heuristics and tools for haptic design.
For example, Ternes and MacLean proposed and empirically validate a series of heuristics for rhythm design~\cite{ternes2008designing} finding that note length and unevenness are key characteristics for discriminability. 
Similarly, Israr et al. introduce a library of haptic \emph{vocabulary}, or mappings between linguistic and haptic patterns and, through VizBiz~\cite{seifi2015vibviz}, Seifi et al. taxonomize haptic characteristics and expose it via an interactive tool for end-user customization.

Despite this work, and a long-running recognition that design guidelines can spur the development of new haptic interfaces~\cite{maclean2008foundations}, much current-day haptic design remains ad hoc.
Designers often approach their respective problems through an iterative approach and by testing each iteration (e.g., as described by the authors of ActiVibe~\cite{cauchard2016activibe} or HaNS~\cite{tam2013design}).
Where current haptic guidelines have been used, designers have found them wanting.
For example, Prasad et al.~\cite{prasad2014designing} use waveform and spatio-temporal patterns to communicate verb phrases through haptic models, and their design choices are informed by the heuristics described above. 
However, on evaluating these designs, haptic performance fared poorly in comparison to auditory performance\,---\,these performances differences are not well-captured by existing design guidelines.

We believe that this gap is due to a lack of work around Chan's fourth factor: how well the saliency of different haptic characteristics persists under cognitive load.
Most empirical work around haptics has had participants attend primarily to the haptic signal, which does not mimic the real-world use cases of haptics as transparent interfaces~\cite{maclean2008foundations}.
When haptic interfaces have been studied in situ~\cite{tam2013design}, it has not followed the systematic approach of empirical studies.
Thus, it has been difficult to refine existing design heuristics and guidelines to reflect the cognitive performance of different haptic characteristics.

Our work begins to address this gap. 
We isolate five haptic properties from Brewster et al.'s taxonomy~\cite{brewster2004tactons}\,---\,amplitude,waveform, duration, rhythm, and spatio-temporal pattern\,---\,and have participants perform a delayed digit recall task (the \emph{1-back}) designed to emulate the auditory and memory load of daily tasks~\cite{mehler2011agelab}.
We measure participants' performance and error rate on both tasks, and gather qualitative and quantitative preferences via the NASA-TLX survey~\cite{hart2006nasa, hart1988development}.
This setup is inspired by work in the automotive industry examining the cognitive impact of car interfaces on drivers~\cite{mahr2012contre} as well as work studying the impact of multi-modal interfaces on cognitive load~\cite{leung2007evaluation}.
Our goal is similarly inspired by work on graphical perception in the data visualization literature~\cite{cleveland1984graphical} which has developed an effectiveness ordering of visual channels (e.g., position, color, size).

%% file: sections/device.tex
\section{Haptic Device Prototype}

The haptic device consists of four Linear Resonant Actuators (LRA). Each actuator is driven by a motor driver (DRV2605L) that has a preset library of over 100 waveforms and amplitudes. Each motor driver is connected to the Bluetooth-enabled Arduino board via a multiplexer for individual control. The LRAs are chosen instead of the Eccentric Rotating Mass motors due to their robustness, the consistency of the vibration pattern they produce and their efficiency. 

\begin{figure}[t]
    \centering
    \includegraphics[width=8.5cm]{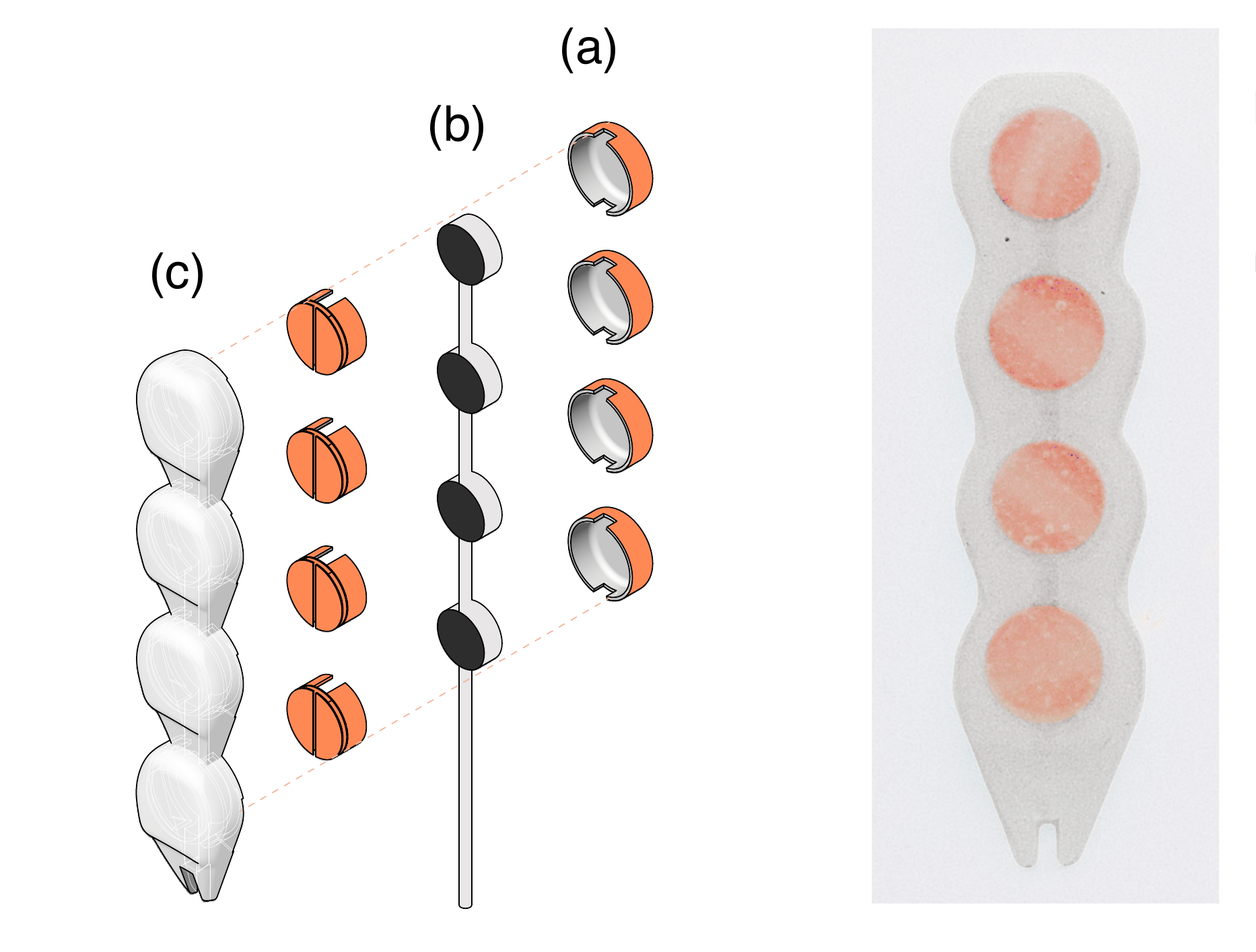}
    \caption{Left: An assembly diagram of the haptic device. (a) actuator housing (b) Four LRAs (Linear Resonant Actuators) (c) silicone casing; Right: Photo of the device prototype.}
    \label{fig:exploded-axon}
\end{figure}

Four motors are arranged in a line to create the spatio-temporal pattern experimental condition (explained in the next section). Each actuator has a 3D-printed housing and is embedded in a high performance, skin-safe silicone rubber casing (see figure \ref{fig:exploded-axon}). The silicone casing conforms to the user’s body. This design allows for maximum skin contact with the actuators and the silicone provides the flexibility needed for the device to maintain full contact with the skin. The device is secured to the participant's forearm using medical-grade tape.

\begin{figure}[t]
    \centering
    \includegraphics[width=8.5cm]{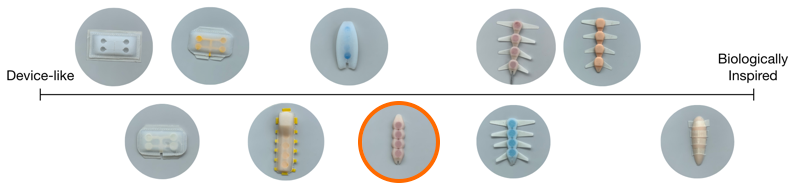}
    \caption{Design Typologies: Our design spectrum spanned from more device-like designs to more biologically-inspired designs. The final design was in the middle of this spectrum.}
    \label{fig:typologies}
\end{figure}

A number of physical forms were explored for the design of the silicone casing, ranging from more device-like to biologically inspired forms. The final design used in this experiment was in between the design spectrum of the two typologies (see figure \ref{fig:typologies}). Additionally, the design iterations explored a 2x2 arrangement of the actuators as well as a linear arrangement. The linear arrangement was selected because it allowed for exploration of multiple spatio-temporal patterns and sequences. 

%% file: sections/method.tex
\section{Experimental Design}

To determine the effectiveness of haptic design while under cognitive load (the fourth goal identified by Chan et al.~\cite{chan2008designing}), we conducted a within-subjects laboratory study with five conditions. 
Participants performed two tasks simultaneously: the \emph{1-back} delayed response task, and a haptic detection task where the gradient of the signal was encoded using one of five haptic properties identified by Brewster et al.~\cite{brewster2004tactons}.

\subsection{1-back Task}

The 1-back task is a delayed digit recall task developed by the MIT Agelab~\cite{mehler2011agelab}.
The participant is presented with a random sequence of auditory stimuli (single digits 0\,--\,9) and are required to respond with the next-to-last stimuli presented. 
This task design approximates the type of auditory and memory load that is induced in daily tasks(e.g., having a phone call or a conversation). 
We measured participants' error rate and recorded their voice to later identify their response time.
Each experimental condition comprised 60 digits, with the first 10 digits used to establish a participant's baseline performance before the secondary task was introduced.

\subsection{Experimental Conditions: Haptic Gradient Detection Task}

For the secondary task, participants were asked to identify the gradient of the haptic signal using their dominant thumb (i.e., thumbs up if they perceive the signal to be increasing, and down for decreasing).
The signal was encoded using only one of five properties drawn from Brewster et al.~\cite{brewster2004tactons}.
We were unable to study frequency due to limitations with the actuators we used in our prototype device. 
We further eliminated body location based on feedback from pilot study participants described in the subsequent subsection.
To reduce confounds, each condition was optimized using existing just noticeable difference (JND) guidelines, and was further refined through piloting to ensure discernibility. 
Additionally, during an initial training phase, the amplitude was tuned to ensure the lowest vibration can be felt by the participant. 
We used an 80\% amplitude base wave as the basis for all signals, and based on the participant sensitivity, the lowest threshold was set to either 40\% or 60\%. 

\begin{figure}[t]
    \centering
    \includegraphics[width=8.5cm]{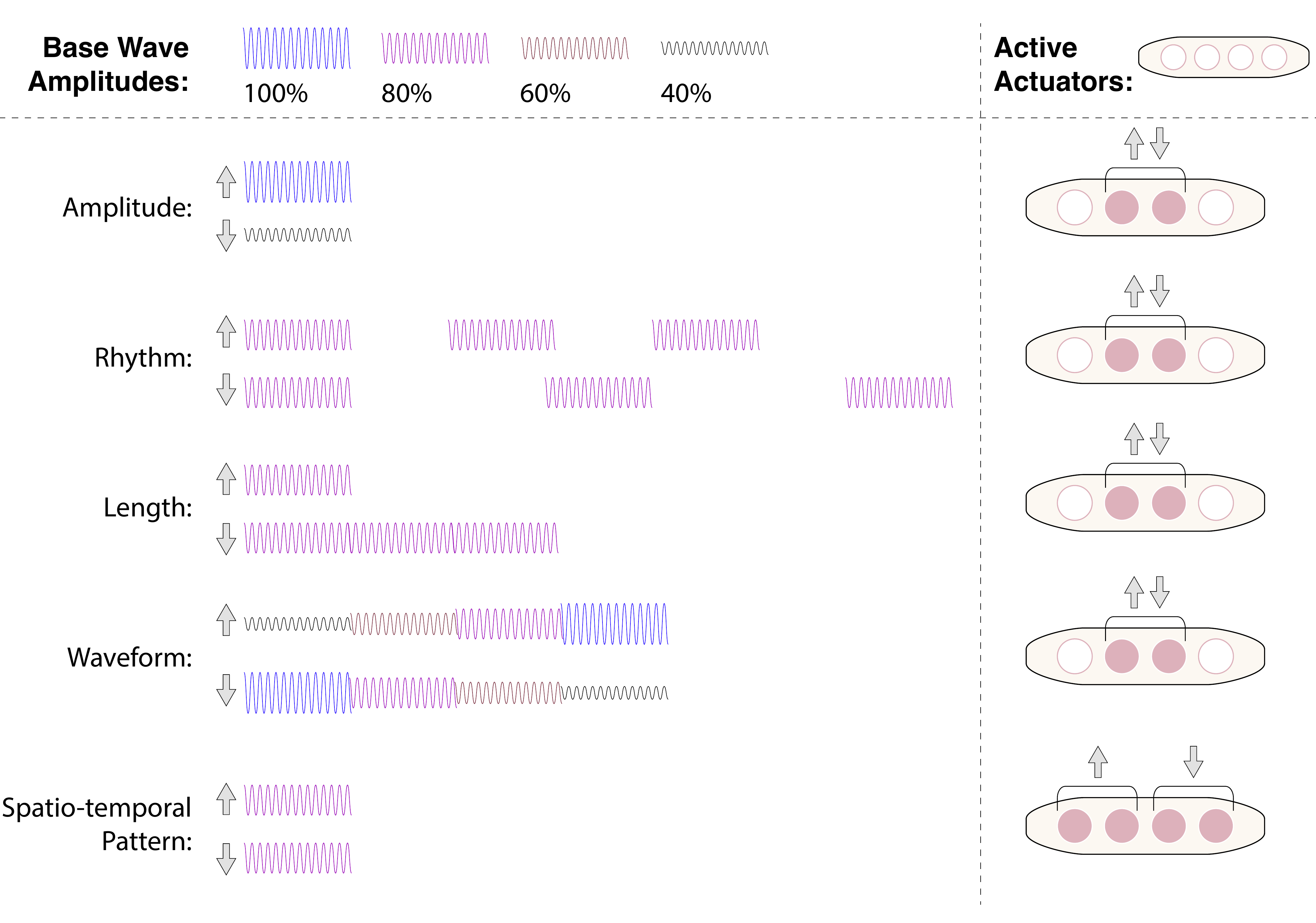}
    \caption{The five experimental conditions of our study. The top row displays the base waves we used, derived from the DRV2605 built-in haptic library. The left-hand side shows how these base waves encode an \emph{up} or \emph{down} signal for each experimental condition (note, the lowest amplitude is set to 40\% or 60\% based on each participant's detection threshold). The right-hand side shows which actuators are active for the condition.}
    \label{fig:conditions}
\end{figure}

The five experimental conditions are as follows (see Fig.\ref{fig:conditions}): 

\begin{enumerate}
    \item \textbf{Amplitude}, or varying the \emph{intensity} of stimulation. In our design the \emph{up} signal was stronger (100\% amplitude) than the \emph{down} signal (40\% or 60\% based on participant sensitivity).
    
    \item \textbf{Rhythm}, or varying how pulses are grouped, or the spacing between pulses. In our design, we used 3 pulses for both conditions. The \emph{up} signal pulsed more quickly with a 500ms interval, and the \emph{down} signal pulsed more slowly with a 1s interval between pulses.
    
    \item \textbf{Duration}, or varying the \emph{length} of one pulse. In our design the \emph{up} signal was the same length as the base wave, and the \emph{down} signal was 3x longer than the base wave.
    
    \item \textbf{Waveform}, or varying the \emph{shape} of the wave. In our design the \emph{up} signal was a ramp going up and the \emph{down} signal was a ramp going down (low threshold set to 40\% or 60\%).
    
    \item \textbf{Spatio-temporal Pattern}, or changes in active actuators over time. As this dimension presents a rich and continuous design space, we picked two alternative designs for simplicity. Our haptic device has four actuators and in all the above conditions the middle two actuators are active. Under this condition, we used the same wave for both up and down signals but varied the set of active actuators\,---\,the first two actuators are active for \emph{up} and the second two for \emph{down}.
\end{enumerate}

\subsection{Pilot Study}

We arrived at our experimental design after conducting a series of pilot studies.
Our pilot design focused on testing two haptic signal designs (encoding gradient as either the waveform or a combination of duration and rhythm) along two body locations (the forearm and upper back) and two spatio-temporal patterns (pattern or no pattern) for a total of 8 experimental conditions.
We installed a haptic device on both body parts at the start of a study session, which was was broken into two phases: in the first phase, participants felt one of the two signal designs on both body parts and with both types of spatio-temporal patterns; then, after a 5 minute break, the study resumed with the second signal design.
Participants were tasked with both the \emph{1-back} and gradient detection task, and we measured their error rate. 
We conducted the pilot with 6 participants, who were compensated \$10 for their time.

Piloting helped us refine our experimental design.
We decided to measure response times on the 1-back task, as error rates alone did not reflect variations in performance observed by researchers.
We chose to additionally administer the NASA-TLX survey~\cite{hart2006nasa, hart1988development} to capture participant preferences.
Perhaps most dramatically, we scoped the final design down to studying each haptic property in isolation.
We eliminated the body part conditions as participants universally disliked the neck position for its inconvenience.  
Additionally, all participants performed better with the second haptic signal design but it was difficult to ascertain why as it entangled two properties (duration and rhythm).
Finally, isolating haptic properties as experimental conditions affords a more uniform design: spatio-temporal patterns are now just one condition, rather than a component studied in conjunction with other properties.

\subsection{Procedure}

\begin{figure}[t]
    \centering
    \includegraphics[width=8cm]{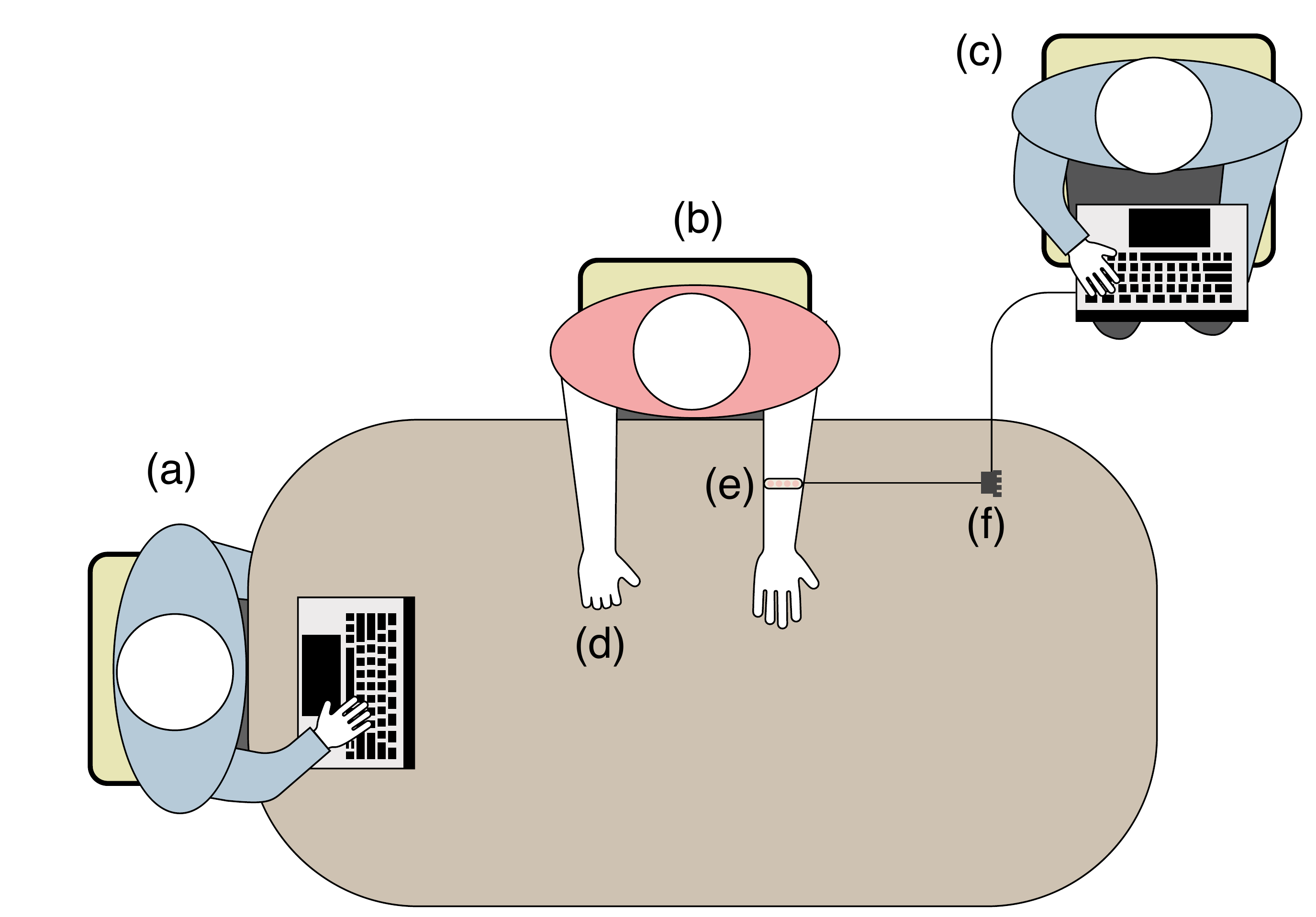}
    \caption{The study setup: (a) one researcher conducted the 1-back task, and recorded errors from (b) the participant; (c) another researcher provided the haptic stimulus and recorded errors to the secondary task; (d) the participant used their dominant thumb to indicate a signal \emph{up} or \emph{down}; (e) the haptic device was worn on the participant's non-dominant forearm, and was (f) connected to the researcher's laptop.}
    \label{fig:setup}
\end{figure}

16 participants (9 female) were recruited via departmental mailing lists and through word-of-mouth. 
They ranged from 21–32 years of age, and received a \$15 gift card as compensation.
The study was conducted in a private meeting room on the university campus. 
In addition to the participant, two researchers were also present in the room (Fig.~\ref{fig:setup}) to control and record performance on each of the two tasks.


The study took approximately 60 minutes to complete.
Each Participant first read and signed a consent form, and researchers provided an overview of the study and explained its goals.
The 1-back task was explained, and the participant had the opportunity to practice the task.
Once the participant felt ready, researchers installed the haptic device on participants' non-dominant hand, and explained the haptic task.
The participants was given a demo of the \emph{up} and \emph{down} signal variants for the first condition, and was instructed to use his dominant thumb to indicate the gradient of the haptic signal.
Each participant received 10 random trials of the up or down signal, to practice interpreting the haptic signal, and indicating its gradient.
If participants were unsure of the gradient, they were instructed to leave their thumb in the horizontal position.

Once training was complete, and once participant was ready, he began to perform the two tasks simultaneously.
For each condition, participants received the first 10 digits (of 60 total digits) of the 1-back task without any haptic signal to establish a baseline.
We determined performance by measuring participants' response time and error rate on each task.

At the end of each condition, participants completed a short survey to determine the interpretability of the signal. 
This survey also contained 5 questions from the NASA-TLX~\cite{hart2006nasa, hart1988development} measured on a 5-point Likert scale: (1) How mentally demanding was the task? (2) How hurried or rushed was the pace of the task? (3) How successful were you in accomplishing what you were asked to do? (4) How hard did you have to work to accomplish your level of performance? (5) How insecure, discouraged, irritated, stressed, and annoyed were you? Finally, participants were asked to provide any open-ended feedback about the condition's haptic signal design.

At the conclusion of the study, participants completed an exit survey.
The survey collected demographic information including their age, gender, ethnicity, and native language\,---\,native language was asked to determine the cognitive load imposed by the 1-back task on non-native speakers.
The survey also asked participants to rank the different conditions based on difficulty in the training and test phases, comment on their current or past use of wearable devices, their preference with regards to body placement of the device, and any comments about their overall experience.


%% file: sections/results.tex
\section{Results}

We used non-parametric statistical methods throughout the analysis to account for the skew in the data and the fact that some responses were ordinal (e.g. Likert Scales). 
When comparing time or performance scores across conditions, we used the Quade test, which is a generalization of the two-sample signed-rank test. Specifically we used R's quade.test from the stats package. 
For all other pairwise comparisons, we used the Wilcoxon test for data coming from the same participant (with R’s wilcox.test function with argument ‘paired’ set to ‘True’)  and the Mann-Whitney test for independent samples (R’s wilcox.test function with argument ‘paired’ set to ‘False’). 
For each raw p-value reported in the analysis, we also provided an alternative adjusted p-value computed via the Holm method.


\subsection{Response Time Comparisons}

\begin{figure}[b]
    \centering
    \includegraphics[width=8.5cm]{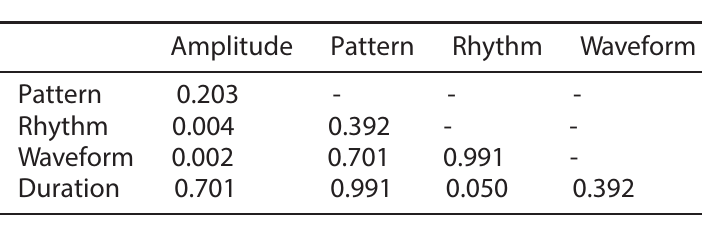}
    \caption{P-value Summary from Post hoc Pairwise Comparisons Using Wilcoxon Signed Rank Test with Holm Adjustment.}
    \label{fig:Kruskal_Time_Table}
\end{figure}

We observed that participants were faster in \textit{Amplitude}, followed by \textit{Duration}, Spatio-Temporal \textit{Pattern}, \textit{Rhythm} and \textit{Waveform}. 
Following a statistically significant Quade test (Quade F = 6.7019, num df = 4, denom df = 60, p-value = 0.0001582), a post hoc pairwise comparison with the Wilcoxon test indicated that \textit{Amplitude} median time was significantly shorter compared to both \textit{Rhythm} (V = 6, p-value = 0.0004272, Holm adjusted p-value = 0.0038) and \textit{Waveform} (V = 3, p-value = 0.0001526, Holm adjusted p-value = 0.0015) (results figure in Table~\ref{fig:Kruskal_Time_Table}). 

\begin{figure}[t]
    \centering
    \includegraphics[width=8.5cm]{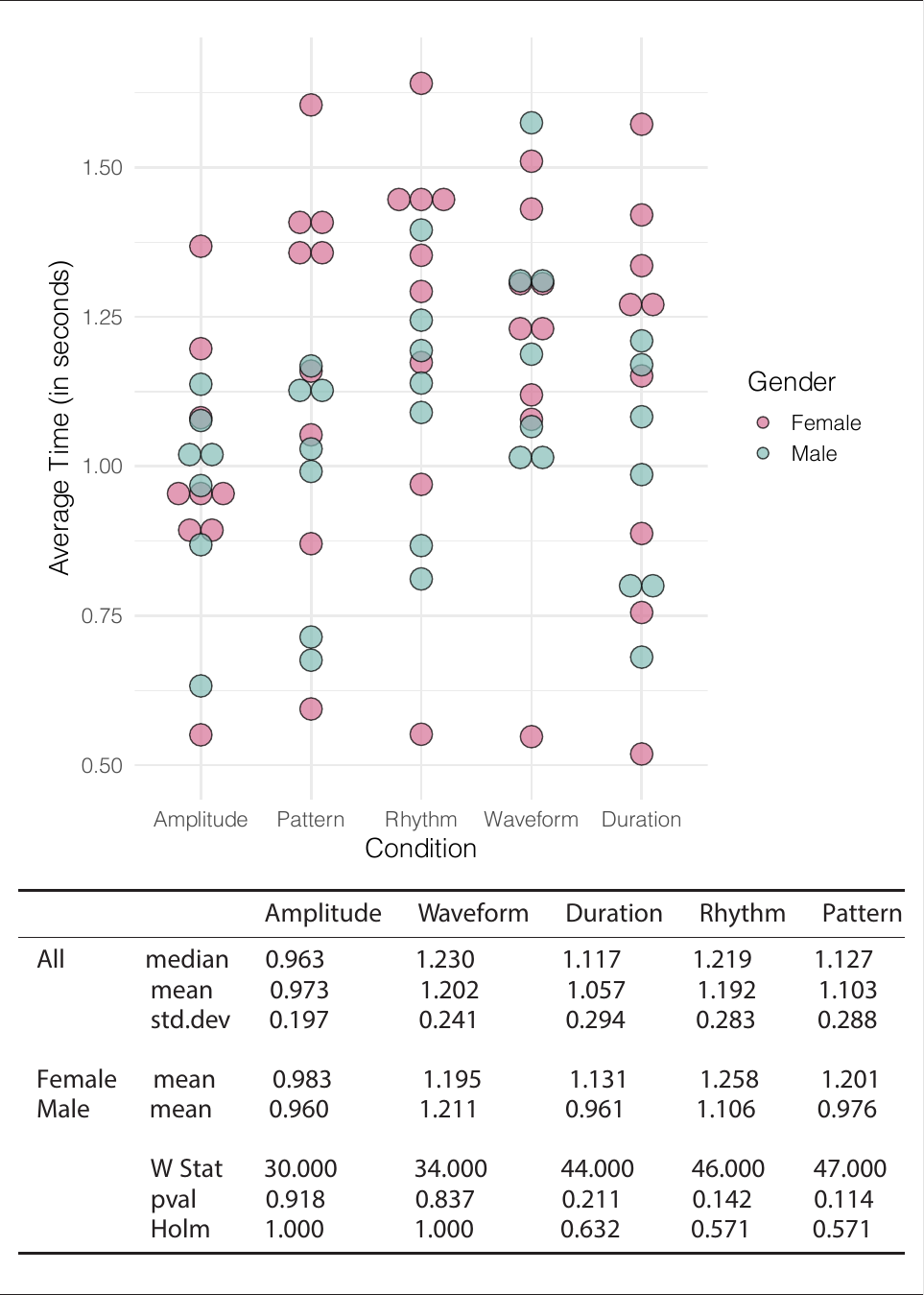}
    \caption{Response Time Differences by Gender across Conditions.}
    \label{fig:Time_Differences}
\end{figure}


We also found that \textit{Rhythm} was significantly longer than \textit{Duration} (V = 17, p-value = 0.006287, Holm adjusted p-value = 0.0503); however, this effect was attenuated after Holm’s adjustment in the pairwise comparison, it remained statistically significant.
Overall, we visually observe that female participants had longer times but also more widely spread distributions compared to males (see figure~\ref{fig:Time_Differences}).

\subsection{Performance in the 1-back Task across Conditions}

\begin{figure}[t]
    \centering
    \includegraphics[width=8.5cm]{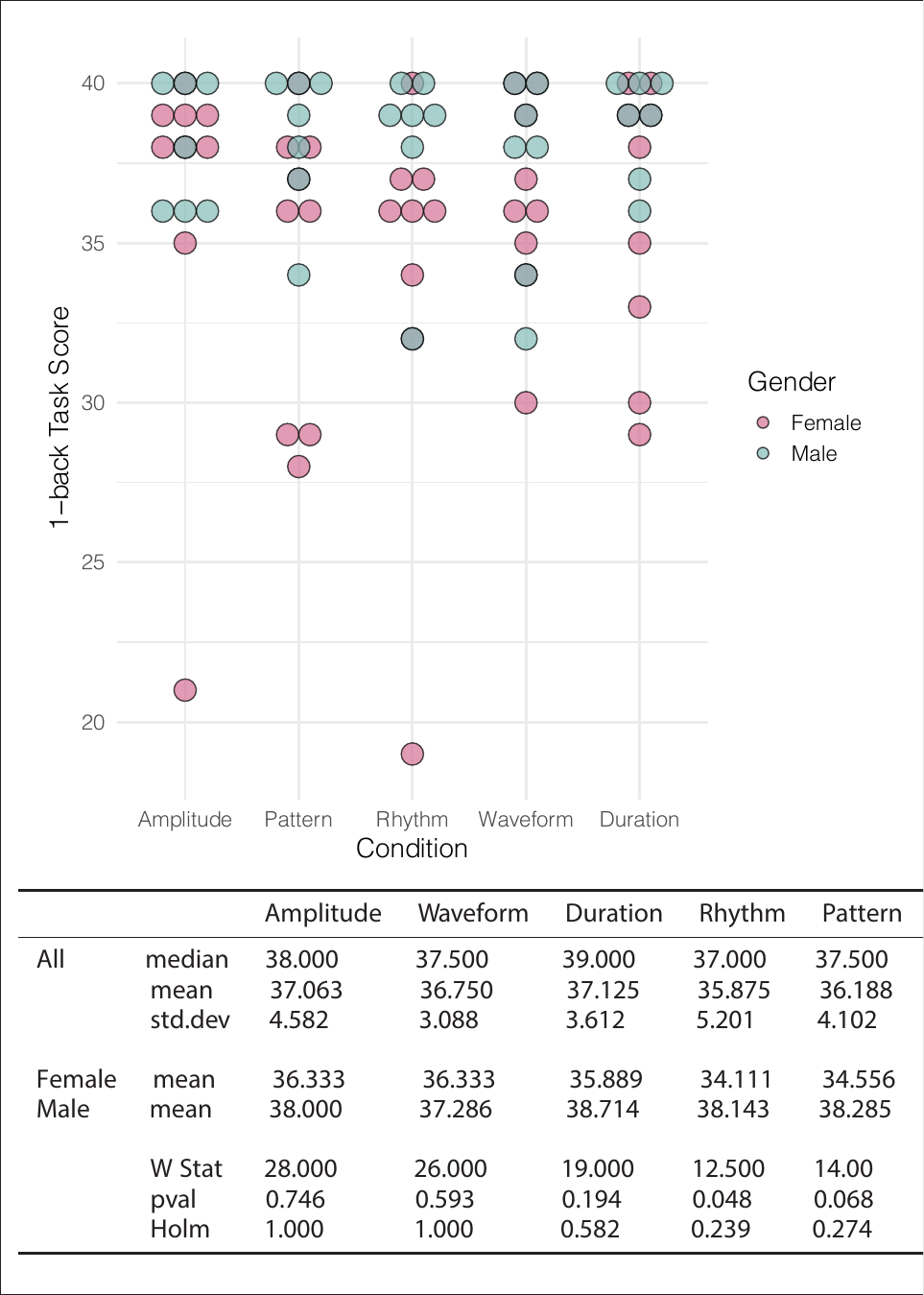}
    \caption{1-back Task Scores.}
    \label{fig:1_back_Scores}
\end{figure}

Although we did not obtain statistically significant differences amongst conditions (see figure \ref{fig:1_back_Scores}). We did observe a visual trend in increased performance at the 1-back task for \textit{Amplitude} and \textit{Duration} (see figure \ref{fig:1_back_Scores}). We also observed that performance in \textit{Rhythm} was the lowest. When breaking down results by Gender, we found that males scored significantly higher in \textit{Rhythm} (W = 12.5, p-value = 0.048) and we almost found a significant difference in \textit{Spatio-Temporal Pattern} (W = 14, p-value = 0.068). While the Holm correction attenuates these effects, we see in the raw data that in contrast to female participants, all males consistently score above thirty points.

\subsection{Performance in the Haptic Gradient Detection Task}

\begin{figure}[t]
    \centering
    \includegraphics[width=8.5cm]{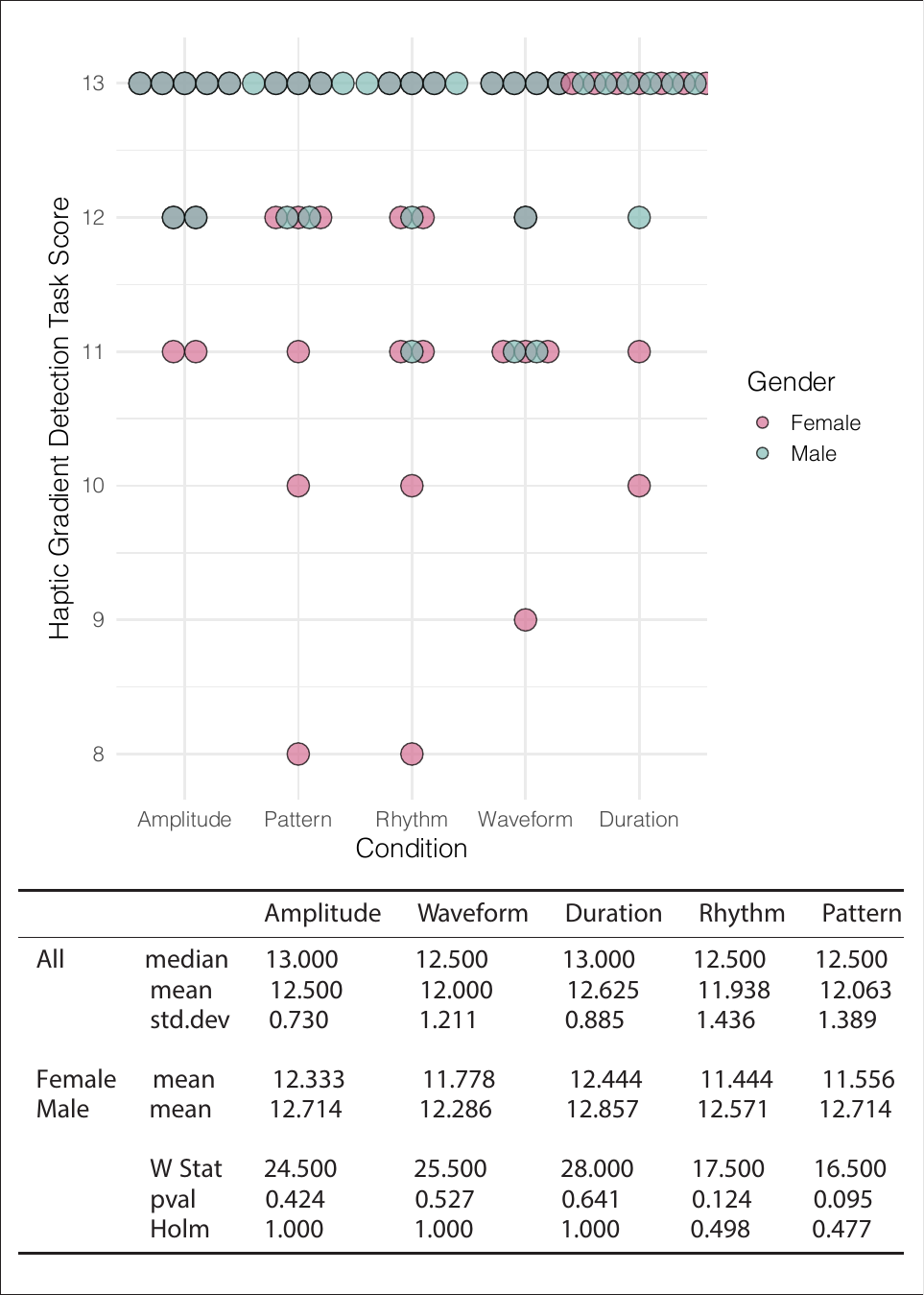}
    \caption{Haptic Gradient Detection Scores.}
    \label{fig:Haptic_Scores}
\end{figure}

While the Quade analysis between conditions did not yield statistically significant results, we observed higher scores during both \textit{Amplitude} and \textit{Duration} which interestingly yield very similar distributions (see figure \ref{fig:Haptic_Scores}). Similarly to the scores from the 1-back task, we also see a decrease in performance in \textit{Rhythm} for both male and female participants. After conducting a more in-depth comparison between males and females showed a difference nearing significance in \textit{Spatio-Temporal Pattern} (W = 16.5, p-value = 0.095). While the Holm correction further attenuates these effects, upon visual inspection, we can see that scores from male participants tend to aggregate towards higher values on the scale (see figure \ref{fig:Haptic_Scores}).

\subsection{NASA-TLX Survey Responses}

\begin{figure*}[t]
    \centering
    \includegraphics[width=18cm]{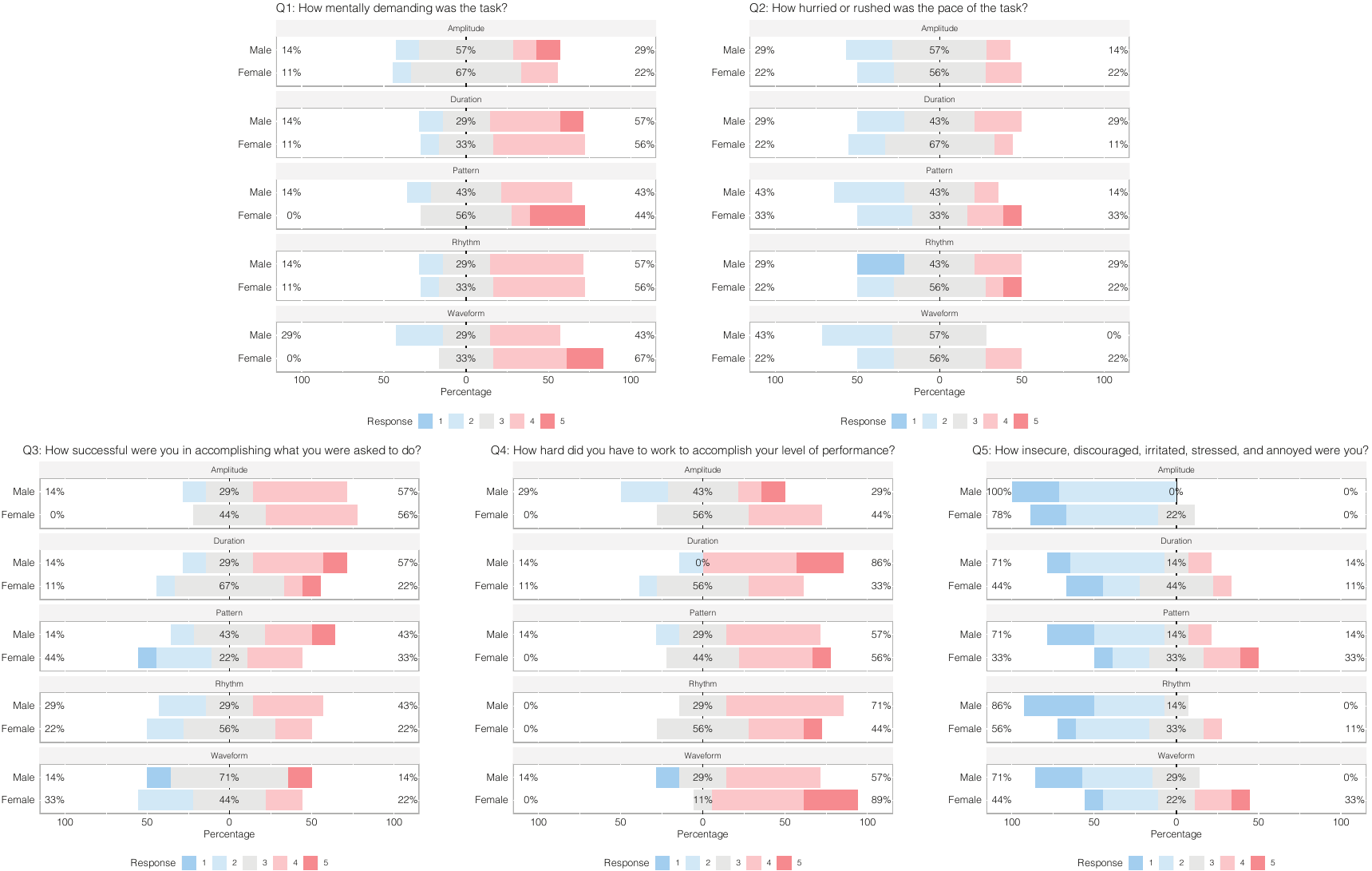}
    \caption{NASA-TLX Responses by Gender across Conditions}
    \label{fig:NASA-TLX Results}
\end{figure*}

A subset of questions from the NASA-TLX allowed us to learn more about each subject’s perceived workload and performance for each condition (see figure \ref{fig:NASA-TLX Results}). We observed that overall participants scored \textit{Amplitude} as the least demanding condition (Question 1) but also the most successful (Question 3). Interestingly, this is consistent with both time and performance data. Besides, \textit{Amplitude} was the task were participant reported having worked the least hard to achieve their level of performance (Question 4) and being the least insecure (Question 5). On the other hand, \textit{Waveform} and \textit{Spatio-Temporal Pattern} ranked overall highest in mental demand (Question 1). \textit{Waveform} was, on average, ranked as the most difficult (Question 4) as well as least successful (Question 3). 
When we took a closer look at differences between genders, we observed additional trends. In all conditions except \textit{Amplitude}, female participants rated themselves as least successful as male participants did. Female participants also rated \textit{Waveform} as the most mentally demanding (Question 1) and most difficult condition (Question 3). Male participants rated \textit{Duration} on average as the most mentally demanding (Question 1) and most difficult condition out of the five (Question 3). Lastly, we observed that female participants rated themselves consistently higher than their male counterparts across conditions as more insecure, discouraged, stressed, and annoyed (Question 5). 

\begin{figure}[p]
    \centering
    \includegraphics[width=8.5cm]{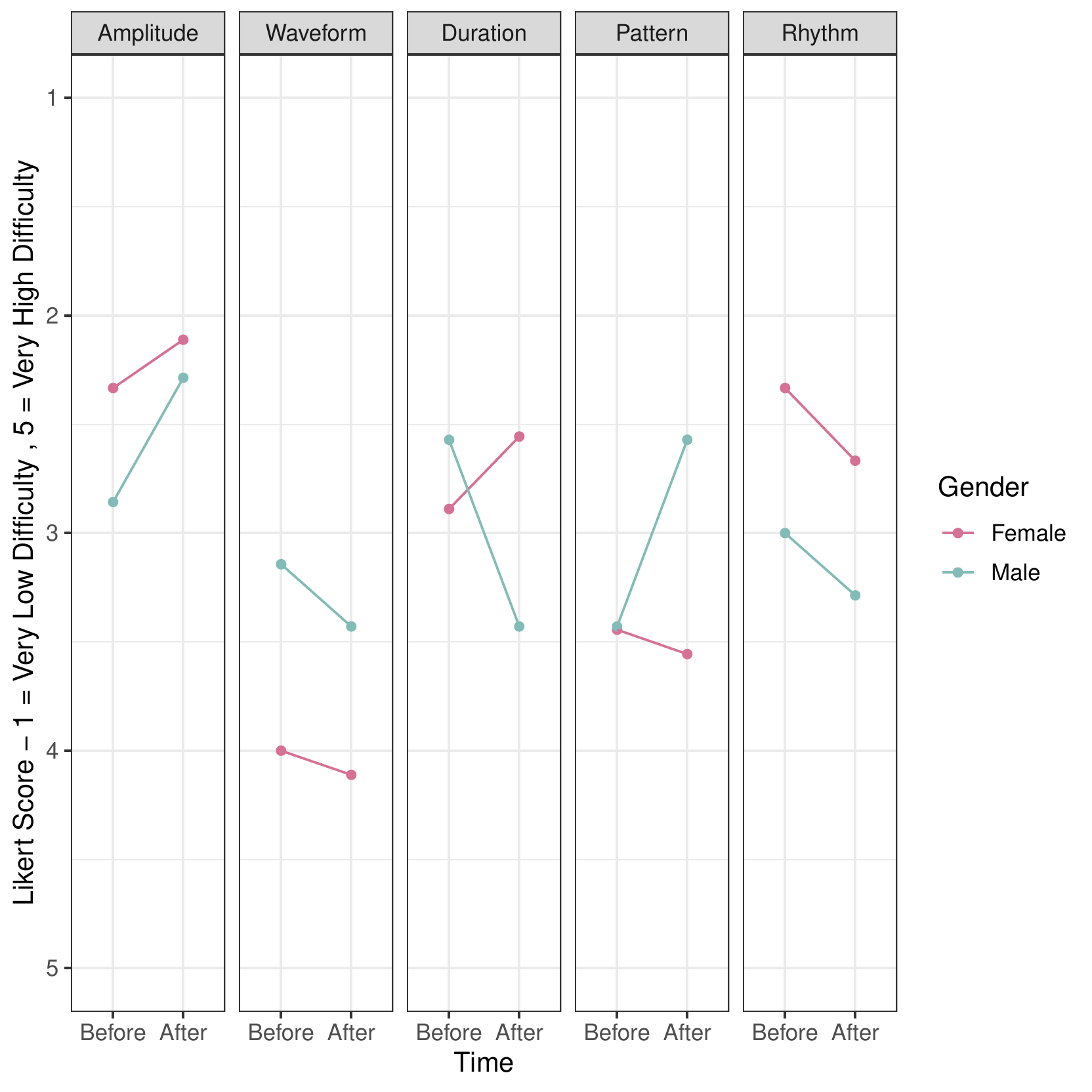}
    \caption{Changes in participant preference ranking of experimental conditions before (during the trial period of the haptic gradient detection task) and after (during the combined haptic gradient detection task and the primary task)}
    \label{fig:before_after_difficulty}
\end{figure}
\begin{figure}[p]
    \centering
    \includegraphics[width=8.5cm]{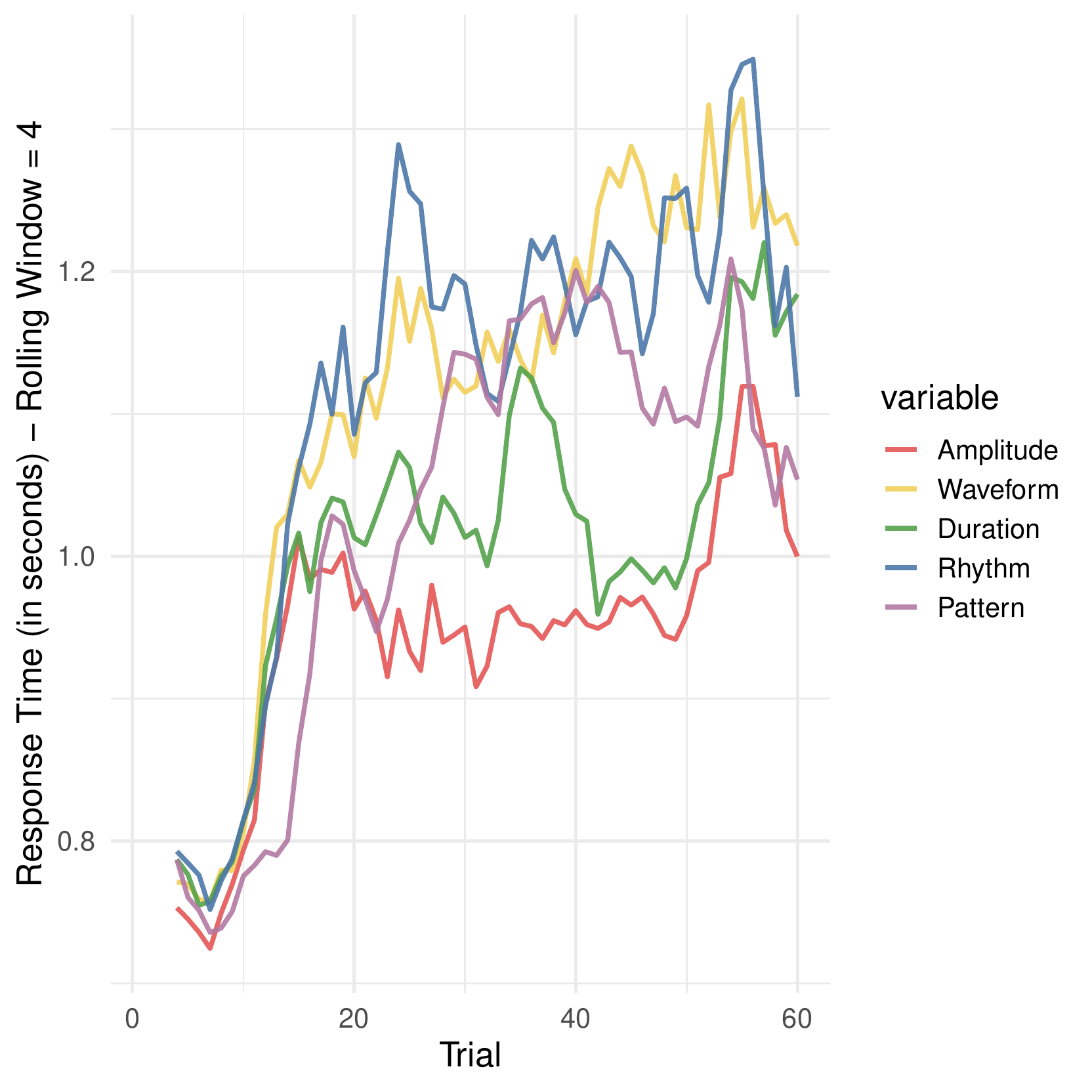}
    \caption{Response Times across Trials.}
    \label{fig:rolling_time}
\end{figure}

We asked all participants to rank the five conditions based on how easily they were able to discern haptic patterns before and after combining it to the 1-back task with ‘1’ being the best and ‘5’ the worst condition. We, subsequently, averaged those responses and grouped them by gender.
While the sample size was relatively small, and the statistical tests were not significant, we were able to observe several trends (see figure \ref{fig:before_after_difficulty}). Both male and female participants ranked \textit{Waveform} and \textit{Rhythm} as more difficult and \textit{Amplitude} as less difficult when the 1-back task was added. Interestingly, for \textit{Duration} and \textit{Spatio-Temporal Pattern} opinions shift between genders. Opposite to females, males ranked \textit{Duration} as harder and \textit{Spatio-Temporal Pattern} as easier. These results are consistent with the NASA-TLX Survey responses as well as with the time and performance scores described in previous sections. 

\subsection{Response Time Trends across Trials}

We computed rolling averages to uncover patterns relative to the passage of time within each condition. A window of four, allowed us to smooth the lines enough so that trends become more prominent while preserving most of the original structure in the data (see figure \ref{fig:rolling_time}). A steep slope becomes apparent as participants transition from the training phase (Trials 1 to 10) to the experimental phase, where they become exposed to the haptic signal. Strikingly, response time remains relatively stable in \textit{Amplitude} while it increases in \textit{Waveform} and \textit{Rhythm} which suggests that there is no habituation for these two conditions. Lastly, we observe an increase in response time across all five conditions after the fiftieth trial. 

\subsection{Summary}
The three data-points collected from the experiment were the response time data from the n-back task, the performance score on the n-back task, and the performance score on the up/down Haptic Gradient Detection task. We consider the response time in the n-back task to be a good indicator of cognitive load imposed on the participant by the haptic condition. 

Results from all participants showed that the response time for \textit{Amplitude} was significantly shorter than \textit{Rhythm} and \textit{Waveform}. \textit{Amplitude} was followed by \textit{Duration} which was significantly shorter than \textit{Rhythm} as well. Although we did not find any statistically significant results on the performance score on the n-back task and on the haptic task, we observed a visual trend whereby participants performed best on both tasks with the \textit{Amplitude} and \textit{Duration} condition. Interestingly, the participant preference on the NASA-TLX survey validated this finding. Overall, participants found \textit{Amplitude} to be the condition ranked as least demanding and most successful.

On the other hand, while not statistically significant, our results revealed that \textit{Waveform} followed by \textit{Rhythm} were the two conditions the participants took the longest to complete. We also found the lowest scores in \textit{Rhythm} and \textit{Spatio-Temporal Pattern} on the n-back task, and \textit{Rhythm}, \textit{Waveform}, and \textit{Spatio-Temporal Pattern} on the haptic task. This finding also aligned with the NASA-TLX results where \textit{Waveform} and \textit{Spatio-Temporal Pattern} ranked as highest mental demand, and \textit{Waveform} was ranked as least successful and most difficult. 

%% file: sections/discussion.tex
\section{Discussion}

In this paper, we begin to systematically study how effectively different haptic properties are able to encode information under cognitive load.
In contrast to prior studies where participants primarily attend to the haptic task, our work takes one step closer to the in situ, real-world situations where haptic feedback is perceived in an ambient fashion.
In addition to revealing differences in response time and error rate, our results also show that the participant preferences change when ranking a haptic property in isolation compared to ranking it in the context of a primary task (see Fig.~\ref{fig:before_after_difficulty}). 
In this section, we will discuss the implications of our findings on the design of haptic interfaces, as well as limitations and future directions.

\subsection{Impact of Signal Length and Time}

Our results showed that the \textit{Amplitude} and \textit{Duration} conditions outperformed \textit{Waveform} and \textit{Ryhthm}. 
One explanation for this result might be differences in the length of the haptic signal. 
Analyzing participants' qualitative feedback revealed that longer signals, and especially the signals that change over time, are more difficult to understand during a simultaneous task as they require shifting attention for a longer period of time. 
On the other hand, signals that are instantly distinguishable and identifiable are easier while performing a simultaneous task.
Although this point was noted by six participants in some way, one participant noted that \textit{Waveform} was their favorite condition because they could \emph{``take their time to process the signal''} and attend to it at their convenience. 
This contrast in opinions may suggest that people employed differing cognitive strategies to complete the two tasks, which impacted the signal design they preferred.

\subsection{Baseline Independent Signals}

The post-experiment survey also revealed that the participants preferred signals which did not rely on a baseline comparison. 
For example, the \textit{Waveform} condition used a ramp design to indicate up or down.
Therefore, the difference between the two signals was inherently built into each signal and did not rely on a comparison between the two signals. 
This can be an important factor to remember when having more than two signals to communicate or when the frequency of information being communicated is low (e.g., one signal every hour).
However, if the difference between the two signals is very clear (e.g., low amplitude is perceptible but extremely weak and high amplitude is very strong), participants do not perform a comparison between the two signals, and thus such an encoding can also be used in less frequent applications.

\subsection{Participant Mental Model}

The last important factor that impacted participant preference was how well the signal matched the user's expectation or mental model. 
Specifically, we found that some participants did not find the \textit{Spatio-temporal pattern} intuitive due to the horizontal orientation of the haptic device on their forearm.
Ironically, we chose this orientation specifically for this experimental condition, hypothesizing that it may be beneficial to the actuators aligned along the cutaneous nerves~\cite{gray1974anatomy}.
This decision resulted in a trade-off for intuitiveness because the vertical orientation may have better matched participants' mental model for up/down. 
One interesting observation is that one participant had a 0\% accuracy on the up/down task in the \textit{Spatio-temporal pattern} condition. 
This participant noted that because the reverse pattern made more sense, they subconsciously adjusted the encoding to mean the reverse of what the experiment had intended. 
They realized this reversal mid-way through, but decided to continue with their preferred encoding. 
As a result, we counted their accuracy on this test as 100\%.

\subsection{Individual Preferences}

Our results also point to several opportunities for personalizing signals based on perception thresholds and individual preferences. 
In the qualitative surveys, we find that subjective ratings and comments were not uniform. 
For example some participants found \textit{Waveform} and \textit{Rhythm} to be the most intuitive whereas others found them the least intuitive. 
Although \textit{Amplitude} and \textit{Duration} were statistically shown to impose a lower cognitive load overall, the impact on performance based on individual preferences should be studied further. 
The minimum and maximum comfortable amplitude for participants was also different. 
Thus, it is important to do a calibration phase to ensure that the highest amplitude does not startle the participant and that the lowest amplitude can be felt but is not too strong. 
This calibration can also become a standard feature in developing wearable haptic devices where the user can set their minimum and maximum desired amplitude and that can be applied to all haptic signals used in the device.

\subsection{Limitations}



Our results indicate that participants' gender or native language significantly affects performance and preference. 
As most of our female participants were also non-native English speakers, it is difficult for us to disentangle these two factors.
The male/native group, preferred \textit{Duration} and \textit{Spatio-temporal patterns} and males nearly significantly performed better on the \textit{Spatio-temporal patterns}. 
Our female/non-native group had, on average, a longer  response time across all conditions which may imply an increased cognitive load compared to the male/native group. 

Given these differences, future work might consider controlling these factors in separate studies.
Additional tweaks to the tasks may also help better isolate these effects.
For instance, future studies may consider using a \emph{shape}-back task (rather than an \emph{n}-back task). 
And, rather than asking participants to repeat the item out loud, a simple interface (whether paper or digital) could be used (e.g., participants could point or click on the item). 
Such an interface may also provide more precise measurements of response times (whereas our current design relied on a digital timer operated by the same researcher).
Of course, such changes would need to be empirically validated outside of the specific context of haptic interfaces to determine how they affect participants' cognitive load in comparison to the n-back task.

Our female/non-native population also reported higher stress ratings on the NASA-TLX survey. 
Future work might consider measuring these values more precisely (e.g., using galvanic skin response as a proxy~\cite{villanueva2016use}) to both determine the degree to which self-reported scores match, but also to assess the impact of stress on cognitive load and the effectiveness of the haptic properties.
Such a design may suggest that particular properties are more desirable in high- or low-stress environments.




\subsection{Future Work}

This paper takes a first step at studying the fourth haptic design factor identified by Chan et al.~\cite{chan2008designing}: how the saliency of haptic signals persists under cognitive workloads.
Our results suggest several promising directions for future work. 

To simplify our experimental design, we chose only a single body location and two spatio-temporal patterns. 
Future research should investigate the degree to which positioning the device for the motor clusters to be supplied by different cutaneous nerves makes a difference in participant performance~\cite{gray1974anatomy}. 
Similarly, future work could investigate the effect of spatio-temporal patterns on various body parts\,---\,in our pilot we found that the perception of signal in the neck is different from the upper back.
Moreover, sensors placed on bones produce different sensations compared to fatty tissues, and some body locations offer the opportunity to include the two in close proximity (e.g., the chest).
In the post-experiment survey, participants reported wrist, hand, forearm, belly, back, neck, and arm as possible locations they would consider using a haptic wearable device. 
Two participants noted that they would have different preferences based on the specific use. These body placements can be explored as alternatives. 

Perhaps most exciting is systematically extending the design of the haptic signal.
To scope our study, we made two simplifying choices.
First, our haptic signal encoded only a single bit of information: whether the signal is going up or down.
Prior work, although it has not studied the cognitive workload impacts, has used haptic feedback to encode discrete (nominal or categorical) information (i.e., tactile icons).
How to use haptic feedback to communicate continuous or quantitative information remains a rich and open question.
Our results already indicate that there may be an effectiveness ordering to the haptic properties (see Fig.~\ref{fig:rolling_time}).
Extending our work to these alternate data types will allow us to develop this ordering more robustly (akin to the effectiveness orderings for visual channels by data type~\cite{munzner2015visualization}).
Second, we chose to study each individual haptic property (e.g., amplitude, waveform, duration, etc.) in isolation.
During our pilot studies, we observed that if multiple properties were used to encode information redundantly (e.g., via amplitude and duration) or different properties were jointly used (e.g., three short pulses for going up vs. one long pulse for going down), it was easier for the participant to understand the data.
Here too, the visual perceptual psychology literature offers some inspiration by studying and categorizing visual channels as integral, separable, or asymmetric~\cite{maceachren2004maps}.



%% file: sections/acknowledgements.tex
\section{Acknowledgments}
The authors would like to thank the participants for their participation in the studies. 

%% file: main.bbl

\begin{thebibliography}{00}


\ifx \showCODEN    \undefined \def \showCODEN     #1{\unskip}     \fi
\ifx \showDOI      \undefined \def \showDOI       #1{{\tt DOI:}\penalty0{#1}\ }
  \fi
\ifx \showISBNx    \undefined \def \showISBNx     #1{\unskip}     \fi
\ifx \showISBNxiii \undefined \def \showISBNxiii  #1{\unskip}     \fi
\ifx \showISSN     \undefined \def \showISSN      #1{\unskip}     \fi
\ifx \showLCCN     \undefined \def \showLCCN      #1{\unskip}     \fi
\ifx \shownote     \undefined \def \shownote      #1{#1}          \fi
\ifx \showarticletitle \undefined \def \showarticletitle #1{#1}   \fi
\ifx \showURL      \undefined \def \showURL       #1{#1}          \fi

\bibitem{6548433}
{M. {Azadi}} {and} {L. {Jones}}. 2013.
\newblock \showarticletitle{Identification of vibrotactile patterns: building
  blocks for tactons}. In {\em 2013 World Haptics Conference (WHC)}. 347--352.
\newblock
\showISSN{null}
\showDOI{%
\url{http://dx.doi.org/10.1109/WHC.2013.6548433}}


\bibitem{azadi2014evaluating}
{Mojtaba Azadi} {and} {Lynette~A Jones}. 2014.
\newblock \showarticletitle{Evaluating vibrotactile dimensions for the design
  of tactons}.
\newblock {\em IEEE transactions on haptics\/} {7}, 1 (2014), 14--23.
\newblock


\bibitem{brewster2004tactons}
{Stephen Brewster} {and} {Lorna~M Brown}. 2004.
\newblock \showarticletitle{Tactons: structured tactile messages for non-visual
  information display}. In {\em Proceedings of the fifth conference on
  Australasian user interface-Volume 28}. Australian Computer Society, Inc.,
  15--23.
\newblock


\bibitem{cauchard2016activibe}
{Jessica~R Cauchard}, {Janette~L Cheng}, {Thomas Pietrzak}, {and} {James~A
  Landay}. 2016.
\newblock \showarticletitle{ActiVibe: design and evaluation of vibrations for
  progress monitoring}. In {\em Proceedings of the 2016 CHI Conference on Human
  Factors in Computing Systems}. 3261--3271.
\newblock


\bibitem{chan2008designing}
{Andrew Chan}, {Karon MacLean}, {and} {Joanna McGrenere}. 2008.
\newblock \showarticletitle{Designing haptic icons to support collaborative
  turn-taking}.
\newblock {\em International Journal of Human-Computer Studies\/} {66}, 5
  (2008), 333--355.
\newblock


\bibitem{cleveland1984graphical}
{William~S Cleveland} {and} {Robert McGill}. 1984.
\newblock \showarticletitle{Graphical perception: Theory, experimentation, and
  application to the development of graphical methods}.
\newblock {\em Journal of the American statistical association\/} {79}, 387
  (1984), 531--554.
\newblock


\bibitem{enriquez2008design}
{Mario~Javier Enriquez}. 2008.
\newblock {\em Design of haptic signals for information communication in
  everyday environments}.
\newblock Ph.D. Dissertation. University of British Columbia.
\newblock


\bibitem{gray1974anatomy}
{Henry Gray} {and} {Charles~Mayo Goss}. 1974.
\newblock \showarticletitle{Anatomy of the human body}.
\newblock {\em American Journal of Physical Medicine \& Rehabilitation\/} {53},
  6 (1974), 293.
\newblock


\bibitem{gunther2001skinscape}
{Eric Eric~Louis Gunther}. 2001.
\newblock {\em Skinscape: A tool for composition in the tactile modality}.
\newblock Ph.D. Dissertation. Massachusetts Institute of Technology.
\newblock


\bibitem{hart2006nasa}
{Sandra~G Hart}. 2006.
\newblock \showarticletitle{NASA-task load index (NASA-TLX); 20 years later}.
  In {\em Proceedings of the human factors and ergonomics society annual
  meeting}, Vol.~50. Sage Publications Sage CA: Los Angeles, CA, 904--908.
\newblock


\bibitem{hart1988development}
{Sandra~G Hart} {and} {Lowell~E Staveland}. 1988.
\newblock \showarticletitle{Development of NASA-TLX (Task Load Index): Results
  of empirical and theoretical research}.
\newblock In {\em Advances in psychology}. Vol.~52. Elsevier, 139--183.
\newblock


\bibitem{israr2006frequency}
{Ali Israr}, {Hong~Z Tan}, {and} {Charlotte~M Reed}. 2006.
\newblock \showarticletitle{Frequency and amplitude discrimination along the
  kinesthetic-cutaneous continuum in the presence of masking stimuli}.
\newblock {\em The Journal of the Acoustical society of America\/} {120}, 5
  (2006), 2789--2800.
\newblock


\bibitem{lee2012evaluation}
{Jaebong Lee} {and} {Seungmoon Choi}. 2012.
\newblock \showarticletitle{Evaluation of vibrotactile pattern design using
  vibrotactile score}. In {\em 2012 IEEE Haptics Symposium (HAPTICS)}. IEEE,
  231--238.
\newblock


\bibitem{leung2007evaluation}
{Rock Leung}, {Karon MacLean}, {Martin~Bue Bertelsen}, {and} {Mayukh
  Saubhasik}. 2007.
\newblock \showarticletitle{Evaluation of haptically augmented touchscreen gui
  elements under cognitive load}. In {\em Proceedings of the 9th international
  conference on Multimodal interfaces}. 374--381.
\newblock


\bibitem{maceachren2004maps}
{Alan~M MacEachren}. 2004.
\newblock {\em How maps work: representation, visualization, and design}.
\newblock Guilford Press.
\newblock


\bibitem{maclean2008foundations}
{Karon~E MacLean}. 2008.
\newblock \showarticletitle{Foundations of transparency in tactile information
  design}.
\newblock {\em IEEE Transactions on Haptics\/} {1}, 2 (2008), 84--95.
\newblock


\bibitem{maclean2009putting}
{Karon~E MacLean}. 2009.
\newblock \showarticletitle{Putting haptics into the ambience}.
\newblock {\em IEEE Transactions on Haptics\/} {2}, 3 (2009), 123--135.
\newblock


\bibitem{mahr2012contre}
{Angela Mahr}, {Michael Feld}, {Mohammad~Mehdi Moniri}, {and} {Rafael Math}.
  2012.
\newblock \showarticletitle{The ConTRe (Continuous Tracking and Reaction) task:
  A flexible approach for assessing driver cognitive workload with high
  sensitivity}. In {\em Adjunct Proceedings of the 4th International Conference
  on Automotive User Interfaces and Interactive Vehicular Applications}. ACM
  New York, NY, 88--91.
\newblock


\bibitem{mehler2011agelab}
{Bruce Mehler}, {Bryan Reimer}, {and} {Jeffery~A Dusek}. 2011.
\newblock \showarticletitle{MIT AgeLab delayed digit recall task (n-back)}.
\newblock {\em Cambridge, MA: Massachusetts Institute of Technology\/} (2011),
  17.
\newblock


\bibitem{meier2015exploring}
{Anita Meier}, {Denys Matthies}, {Bodo Urban}, {and} {Reto Wettach}. 2015.
\newblock \showarticletitle{Exploring Vibrotactile Feedback on the Body and
  Foot for the Purpose of Pedestrian Navigation}.
\newblock
\showDOI{%
\url{http://dx.doi.org/10.1145/2790044.2790051}}


\bibitem{munzner2015visualization}
{T. Munzner}. 2015.
\newblock {\em Visualization Analysis and Design}.
\newblock CRC Press.
\newblock
\showISBNx{9781498759717}
\showURL{%
\url{https://books.google.de/books?id=NfkYCwAAQBAJ}}


\bibitem{pongrac2008vibrotactile}
{Helena Pongrac}. 2008.
\newblock \showarticletitle{Vibrotactile perception: examining the coding of
  vibrations and the just noticeable difference under various conditions}.
\newblock {\em Multimedia systems\/} {13}, 4 (2008), 297--307.
\newblock


\bibitem{prasad2014designing}
{Manoj Prasad}, {Murat Russell}, {and} {Tracy~A Hammond}. 2014.
\newblock \showarticletitle{Designing vibrotactile codes to communicate verb
  phrases}.
\newblock {\em ACM Transactions on Multimedia Computing, Communications, and
  Applications (TOMM)\/} {11}, 1s (2014), 1--21.
\newblock


\bibitem{scott2008comparison}
{JJ Scott} {and} {Robert Gray}. 2008.
\newblock \showarticletitle{A comparison of tactile, visual, and auditory
  warnings for rear-end collision prevention in simulated driving}.
\newblock {\em Human factors\/} {50}, 2 (2008), 264--275.
\newblock


\bibitem{seifi2019personalizing}
{Hasti Seifi}. 2019.
\newblock {\em Personalizing Haptics}.
\newblock Springer.
\newblock


\bibitem{seifi2016exploring}
{Hasti Seifi} {and} {Kent Lyons}. 2016.
\newblock \showarticletitle{Exploring the design space of touch-based
  vibrotactile interactions for smartwatches}. In {\em Proceedings of the 2016
  ACM International Symposium on Wearable Computers}. 156--165.
\newblock


\bibitem{seifi2015vibviz}
{Hasti Seifi}, {Kailun Zhang}, {and} {Karon~E MacLean}. 2015.
\newblock \showarticletitle{VibViz: Organizing, visualizing and navigating
  vibration libraries}. In {\em 2015 IEEE World Haptics Conference (WHC)}.
  IEEE, 254--259.
\newblock


\bibitem{summers1997information}
{Ian~R Summers}, {Philip~G Cooper}, {Paul Wright}, {Denise~A Gratton}, {Peter
  Milnes}, {and} {Brian~H Brown}. 1997.
\newblock \showarticletitle{Information from time-varying vibrotactile
  stimuli}.
\newblock {\em The Journal of the Acoustical Society of America\/} {102}, 6
  (1997), 3686--3696.
\newblock


\bibitem{tam2013design}
{Diane Tam}, {Karon~E MacLean}, {Joanna McGrenere}, {and} {Katherine~J
  Kuchenbecker}. 2013.
\newblock \showarticletitle{The design and field observation of a haptic
  notification system for timing awareness during oral presentations}. In {\em
  Proceedings of the SIGCHI Conference on Human Factors in Computing Systems}.
  1689--1698.
\newblock


\bibitem{ternes2008designing}
{David Ternes} {and} {Karon~E Maclean}. 2008.
\newblock \showarticletitle{Designing large sets of haptic icons with rhythm}.
  In {\em International Conference on Human Haptic Sensing and Touch Enabled
  Computer Applications}. Springer, 199--208.
\newblock


\bibitem{villanueva2016use}
{Idalis Villanueva}, {Maria Valladares}, {and} {Wade Goodridge}. 2016.
\newblock \showarticletitle{Use of galvanic skin responses, salivary
  biomarkers, and self-reports to assess undergraduate student performance
  during a laboratory exam activity}.
\newblock {\em JoVE (Journal of Visualized Experiments)\/} 108 (2016), e53255.
\newblock


\end{thebibliography}
